\documentclass[journal,11pt]{IEEEtran}
\ifCLASSINFOpdf
\else
\fi
\usepackage[tight,footnotesize]{subfigure}

\usepackage{pstricks,graphicx}
\usepackage{amssymb}
\usepackage{amsmath}
\usepackage{setspace}
\usepackage{psfrag}

\newtheorem{theorem}{Theorem}[section]

\newtheorem{proposition}[theorem]{Proposition}

\newcommand{\qed}{\nobreak \ifvmode \relax \else
      \ifdim\lastskip<1.5em \hskip-\lastskip
      \hskip1.5em plus0em minus0.5em \fi \nobreak
      \vrule height0.75em width0.5em depth0.25em\fi}

\begin{document}

\title{Optimal Index Assignment for\\Multiple Description Scalar Quantization}
%
%
%

\author{Guoqiang~Zhang\thanks{Guoqiang Zhang is with the Department of Mediamatics, Delft University
of Technology, Delft, The Netherlands (e-mail: g.zhang-1@tudelft.nl)},
        Janusz~Klejsa\thanks{Janusz Klejsa is with the School of Electrical Engineering, KTH
–- Royal Institute of Technology, Stockholm, Sweden, (e-mail: janusz.klejsa@ee.kth.se)},
        and~W.~Bastiaan~Kleijn\thanks{W. Bastiaan Kleijn is with both the School of Electrical Engineering, KTH
–- Royal Institute of Technology, Stockholm, Sweden and the School of Engineering and Computer
Science, Victoria University of Wellington, New Zealand (e-mail: bastiaan.kleijn@ecs.vuw.ac.nz)}
}

\onecolumn
\doublespacing

\maketitle

\begin{abstract}
We provide a method for designing an optimal index assignment for scalar $K$-description coding. The method stems from a construction of translated scalar lattices, which provides a performance advantage by exploiting a so-called staggered gain. Interestingly, generation of the optimal index assignment is based on a lattice in $K-1$ dimensional space. The use of the $K-1$ dimensional lattice facilitates analytic insight into the performance and eliminates the need for a greedy optimization of the index assignment. It is shown that that the optimal index assignment is not unique. This is illustrated for the two-description case, where a periodic index assignment is selected from possible optimal assignments and described in detail. The new index assignment is applied to design of a $K$-description quantizer, which is found to outperform a reference $K$-description quantizer at high rates. The performance advantage due to the staggered gain increases with increasing redundancy among the descriptions. 
\end{abstract}



%
\IEEEpeerreviewmaketitle

\section{Introduction}
%
%
%
%
\IEEEPARstart{T}HE real-time transmission of multimedia content over
contemporary packet-switched networks generally requires a coding
scheme that can address the effects of packet loss. Multiple-description coding (MDC) creates a plurality of descriptions of a source signal. The descriptions are embedded into packets and transmitted over a lossy network. The descriptions are mutually refinable and any subset of them can be used to reconstruct the source signal. The more descriptions reach the destination the better the reconstruction is. This property naturally requires the introduction of redundancy, which is used to adjust the trade-off among the distortions arising from the possible description loss scenarios.

\par Multiple description quantizers are the most prominent class of MDC schemes, since they are almost directly applicable in many practically relevant source coding scenarios. The first multiple description quantization schemes were proposed by Vaishampayan for a two-description scalar case \cite{Vaishampayan93Multiple, Vaishampayan94MDC}. The scalar schemes were extended to the two-description vector case in \cite{servetto99multiple, vaishampayan01multiple, Goyal2002}. A symmetric $K$-description vector scheme was proposed by {\O}stergaard \emph{et al.} in \cite{Ostergaard06MDC} and extended to even more general asymmetric case in \cite{Ostergaard2010}.
\par A majority of MDC quantization schemes is based on the construction of a fine ``central'' quantizer and $K$ coarser side quantizers. The central quantizer is related to the side quantizers by means of an index assignment mapping, which uniquely maps a quantization index to a tuple of $K$ indices that are associated with the side quantizers. The design of such an index assignment plays a central role in optimizing the performance of MDC.
\par The research on index assignments for MDC is vast. The first practical index assignment schemes were proposed in \cite{Vaishampayan93Multiple}. The design of an efficient index assignment can be formulated as a combinatorial problem \cite{Berger-Wolf02IndexAssignment, Cardinal2004}, a graph optimization \cite{Dumitrescu2007}, a transporation problem in operations research \cite{Liu2009}. If lattice codebooks are used, the design problem of the index assignment can be formulated as a labeling problem, where each lattice point of the central quantizer is mapped to a unique $K$-tuple consisting of points belonging to $K$ respective lattice side quantizers. The labeling problem was solved in \cite{Ostergaard06MDC, Ostergaard2010} by exploiting properties of clean sublattices and using bipartite matching. An efficient index assignment scheme was proposed in \cite{huang06Multiple}, where an auxiliary so-called fractional lattice was introduced significantly simplifying the labeling problem. The fractional lattice was also used in \cite{Liu2009}, where an index assignment based on a transportation model was developed.

\par This paper aims at designing an optimal index assignment for
$K$-channel multiple-description scalar quantization. The proposed index assignment scheme utilizes a so-called staggered gain, which stems from a construction of translated lattices. Several practical multiple description schemes attempt to utilize this gain in a heuristic manner for a two-description scalar case \cite{Tian04MDC_scalar}, a two-description vector case \cite{Tian05MDC} and $K$-description case with a two-stage coding \cite{Samarawickrama2010}. In our work the index assignment exploits the staggered gain for an arbitrary number of descriptions. The staggered gain becomes considerable in the case of high redundancy among the descriptions. However, the gain vanishes as the redundancy decreases and the performance of the proposed index assignment becomes equivalent to existing schemes (e.g. \cite{huang06Multiple}). Nevertheless, the new index assignment is generally advantageous, since the cases, where the redundancy vanishes, are of low practical importance.

\par In the proposed scheme, a central
quantizer and $K$ side quantizers are designed to be translated
$\mathcal{Z}^1$ lattices. A simple mapping function from a central
point to a $K$-tuple of the cartesian product of the $K$ side
quantizers is proposed. The selection of a good $K$-tuple to label a
central point is performed by choosing a point of a translated
$A_{K-1}$ lattice with a short distance from the origin. The index
assignment is shown to be optimal under a common decoding
process.
\par In this work, we use a so-called reference quantizer that resembles the idea of a fractional lattice \cite{huang06Multiple}, which facilitates the design of the index assignment. A difference to work of \cite{huang06Multiple} and \cite{Liu2009} is that the reference quantizer arises from a construction of translated side quantizers, which utilizes the staggered gain. We show, that using the proposed reference quantizer, we only need to label the central points in a reference quantizer cell. Further, the labeling operation becomes straightforward by using the $A_{K-1}$ lattice. Thus, the labeling complexity for the proposed index assignment is reduced compared to that of \cite{huang06Multiple} and \cite{Liu2009}. In addition, the performance for the case of non-vanishing redundancy is improved.
\par The use of the $A_{K-1}$ lattice in designing of the index assignment has a number of practical advantages. It facilitates generation of the index assignment at hand, leads to low operational complexity of an MDC scheme and provides analytic insight into the performance of the obtained multiple description quantizer. We illustrate these properties by describing a regular index assignment for the two-description case in more detail. The index assignment based on the $A_{K-1}$ lattice is applied to the $K$-description scalar quantization. The performance advantage over \cite{huang06Multiple} is demonstrated. The high-rate performance is further evaluated by considering a distortion product \cite{Zhang2011}.

\section{Preliminaries}
\label{sec:preliminaries}
Suppose a source random variable $V$ is to be encoded and transmitted
through $K$ channels. Denote its realization as $v$. It is first quantized to the nearest point
$\lambda_c=\mathcal{Q}(v)$ of a central quantizer $\mathcal{A}_c$. $K$
descriptions of $\lambda_c$ are then produced and transmitted through
$K$ separate channels. Assuming symmetry of the channel conditions, we
consider designing balanced descriptions, where the transmission rate
is the same per channel and the decoding distortion only depends on
the number of received descriptions. The \emph{mean squared error} (MSE)
is taken as the distortion measure.

\begin{figure}[htb]
\centering
\begin{footnotesize}
  \psfrag{1}[c][c]{source}
  \psfrag{2}[c][c]{\footnotesize{des. 1}}
  \psfrag{3}[c][c]{\footnotesize{des. $K$}}
  \psfrag{a}[c][c]{Encoder}
  \psfrag{b}[c][c]{Network}
  \psfrag{c}[c][c]{$\alpha^{-1}$}
  \psfrag{d}[c][c]{averaging}
  \psfrag{e}[c][c]{$D_{(K,K)}$}
  \psfrag{f}[c][c]{$D_{(K,\kappa)}$}
  \psfrag{i}[c][c]{{\footnotesize $\vdots$}}
  \psfrag{g}[c][c]{$\kappa=K$}
  \psfrag{h}[c][c]{$ \kappa< K$}
  \includegraphics[width=70mm]{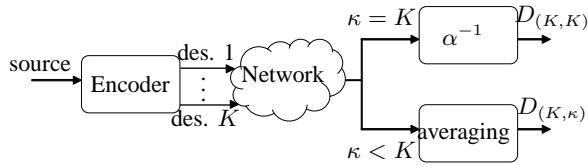}
\end{footnotesize}
\caption{\small{The schematic MDC scheme of the proposed index assignment. The quantity $\kappa$ indicates the number of received descriptions.}} \label{fig:MDC_scheme_illustration}
\end{figure}

Each description is a quantization index describing the associated
side quantizer point. We denote the $K$ side quantizers as $\mathcal{A}_i$,
$i=0,\ldots,K-1$. The
descriptions are produced through an injective labeling function
$\alpha:\mathcal{A}_c\rightarrow\mathcal{A}_0\times
\mathcal{A}_1\ldots \times\mathcal{A}_{K-1}$, expressed as
\begin{equation}
\alpha(\lambda_c)=(\lambda_0,\lambda_1,\ldots,\lambda_{K-1}), \quad \lambda_i\in \mathcal{A}_i\mbox{.} \label{equ:alpha}
\end{equation}
We write each component function of $\alpha$ as $\alpha_i$,
$i=0,\ldots,K-1$. The labeling function usually results in that the central cells associated with each side quantizer point is disjoint. Upon receiving all the descriptions, the central
point $\lambda_c$ is determined uniquely by the inverse mapping
$\alpha^{-1}$. This requires that each $K$-tuple
$(\lambda_0,\lambda_1,\ldots,\lambda_{K-1})$ is used at most once. In
principle, there are $2^K-1$ decoders as there are that many
possible channel states. As the channels are symmetric, we consider
designing a MDC scalar quantizer such that the decoding operation and
the (mean) distortions are only affected by the number of received
descriptions. To achieve this goal, we exploit a common
decoding process \cite{Ostergaard06MDC}. Suppose $\kappa$ $(1\leq
\kappa<K)$ out of $K$ descriptions are received. Considering which
$\kappa$ description are received, there are ${K \choose \kappa}$
different configurations. Let $\mathcal{L}^{(K,\kappa)}$ denote the
set consisting of all the possible configurations. We denote the
$\kappa$-tuple associated with an element $l\in
\mathcal{L}^{(K,\kappa)}$ as $\{\lambda_{l_j},
j=1,\ldots,\kappa\}$. The reconstruction of the source $V$ for some
$l\in \mathcal{L}^{(K,\kappa)}$ is taken as the average of the
received descriptions \cite{Ostergaard06MDC}:
\begin{equation}\hat{V}=\frac{1}{\kappa}\sum_{j=1}^{\kappa} \lambda_{l_j}\mbox{.}\label{equ:X_reconstruction}\end{equation}
Strictly speaking, the estimator $\hat{V}$ in
(\ref{equ:X_reconstruction}) might not be optimal. To achieve optimal
estimation, the central cells that contribute to the element $l$ must
be known, complicating the design problem. Conversely, the use of
(\ref{equ:X_reconstruction}) provides a
good estimate and facilitates the design of index assignment.
Under the averaging operation in (\ref{equ:X_reconstruction}), the
decoding system is essentially simplified to two decoders, see
Fig. \ref{fig:MDC_scheme_illustration}. The quantity $D_{(K,\kappa)}$
denotes the distortion when $\kappa$ descriptions are
received. $D_{(K,K)}$ is referred to as a central distortion, and
$D_{(K,\kappa)}$, $\kappa=1,\ldots,K-1$ as side distortions.

Upon specifying the decoding process, the challenge is to design an
index assignment (specify the labeling function $\alpha$) to achieve
good performance. Optimality and simplicity of the index assignment are
the two main concerns in our work.

\section{Index Assignment}
\label{sec:IA}

We first describe the arrangement of a central and $K$ side
quantizers. Then we formulate optimality of an index assignment. After that, we present the proposed optimal index assignment. We then study the properties of the index assignment. Finally, we consider the index assignment for the two-channel case in the form of an IA matrix.

\subsection{Setup of Central and Side Quantizers}
\label{subsection:quan_structure}
The side quantizers are constrained to be translated $\mathcal{Z}^1$
lattices. For the case of $K > 1$ descriptions, we define the side
quantizers as
\begin{equation}\mathcal{A}_i=\{K\zeta\cdot
  x_i+(2i-K+1)\frac{\zeta}{2}| x_i\in\mathbb{Z}\}, \textrm{ }
  i=0,\ldots,K-1\mbox{,} \label{equ:side_quantizer}\end{equation}
where the scaling parameter $\zeta$ is introduced to adjust the distances between
the side quantizer points. We refer to $x_i$ in (\ref{equ:side_quantizer}) as the
coordinate of $\mathcal{A}_i$. The $K$ side quantizers are arranged so that
$\mathcal{A}_{i+1}$ is obtained by translating
$\mathcal{A}_{i}$ leftward by $\frac{1}{K}$ of the (side) cell
width. Intuitively speaking, the arrangement produces a joint
quantizer that achieves lower distortion than that of each
side quantizer. The performance improvement due to translating side
quantizer to produce a finer joint quantizer is referred to as a
\emph{staggering gain} \cite{Frank-Dayan02MDC}. See also
\cite{Tian04MDC_sequential} and \cite{Tian05MDC_scalar} for MDC
schemes exploiting translated lattices.

The joint quantizer obtained by combining the $K$ side quantizers is
referred to as the \emph{reference} quantizer, denoted as
$\mathcal{A}_r$. The reason for introducing the
reference quantizer is to separate the arrangement of a central
quantizer from the $K$ side quantizers.
The centroids of the reference cells take the form
\begin{equation}
\mathcal{A}_r= \{\zeta\cdot z | z\in \mathbb{Z}\}\mbox{.}
\end{equation}
It can be easily shown that
\begin{equation}\mathcal{A}_r=\frac{1}{K}\sum_{i=0}^{K-1}\mathcal{A}_i\mbox{,}\label{equ:reference_side_relation}\end{equation}
which states that the centroid of any $K$-tuple is a point of $\mathcal{A}_r$. We point out that the fractional lattice introduced in \cite{huang06Multiple} has the same property as (\ref{equ:reference_side_relation}). Specifically, in \cite{huang06Multiple}, the fractional lattice is defined to be a side lattice scaled by $1/K$, and without lattice translation. In our work, the reference quantizer arises naturally from the arrangement of side lattice quantizers.
The reference cell width is $\zeta$ as compared to the cell width
$K\zeta$ of a side quantizer. Thus, each side quantizer cell contains $K$
points of the reference quantizer.

\begin{figure}[htb]
\centering
\begin{footnotesize}
  \psfrag{a}[c][c]{{$\zeta$}}
  \psfrag{b}[c][c]{{$\frac{\zeta}{M}$}}
  \psfrag{c}[c][c]{{side quantizer}}
  \psfrag{d}[c][c]{{reference quantizer}}
  \psfrag{f}[c][c]{{central quantizer}}
  \psfrag{e}[c][c]{$\ldots$}
  \includegraphics[width=70mm]{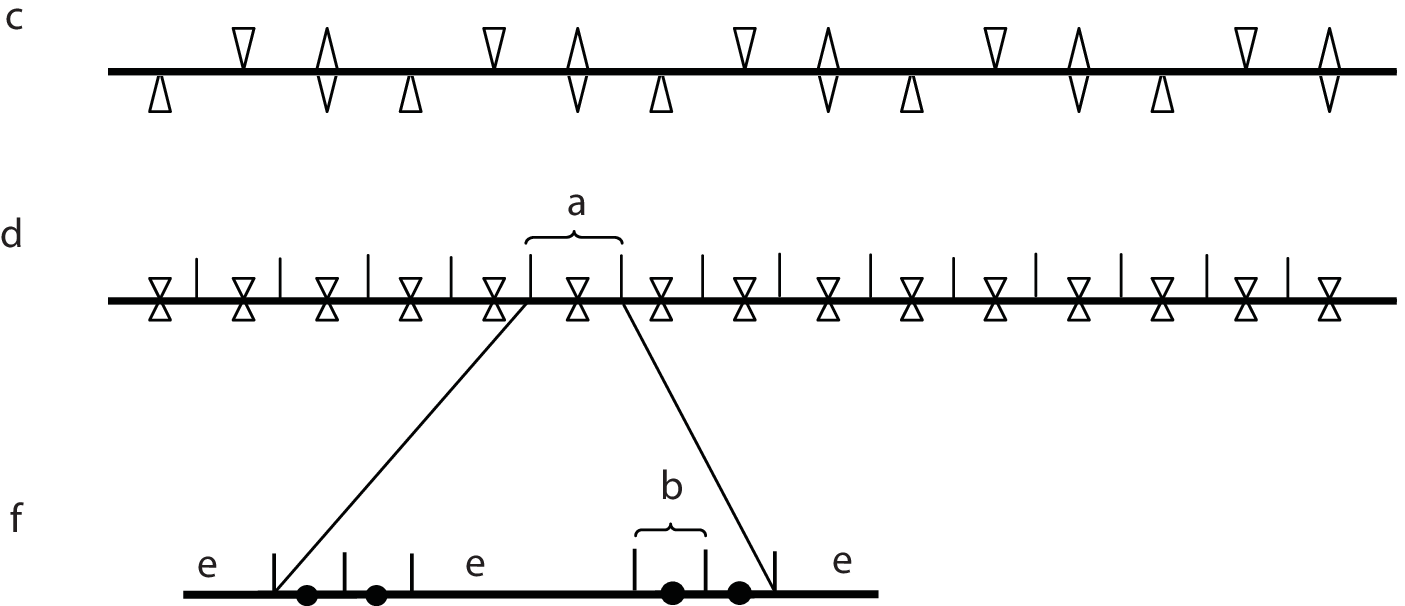}
\end{footnotesize}
\caption{\small{The quantization structure for $K=3$. The points denoted by $\vartriangle$, $\triangledown$ and $\lozenge$ represent the points of the three side quantizers. }} \label{fig:MDSW_QuanStructure}
\end{figure}

Based on the reference quantizer, the central quantizer is defined as
\begin{equation}\mathcal{A}_{c}=\{\frac{\zeta}{M}\cdot
  y+\frac{\zeta}{2M}\cdot\textrm{mod}(M+1,2):y\in
  \mathbb{Z}\}\mbox{,}\label{equ:central_quantizer}\end{equation}
where $M$ is an integer. Thus, the cell width of the central
quantizer is $\frac{\zeta}{M}$. 
The definition guarantees that there are $M$ central points within each
reference quantizer cell and also that the distribution of the points
within each cell is the same. From (\ref{equ:side_quantizer}) and
(\ref{equ:central_quantizer}), it is immediate that there are $KM$
central points within each side quantizer cell. We define $N=KM$ as the redundancy index. The parameter $M$ (or $N$) serves as a trade-off factor between central
distortion and side distortion, which can take any value of
$\mathbb{Z}^{+}$. An example of the quantization structure for $K=3$
is illustrated in $\textrm{Fig. \ref{fig:MDSW_QuanStructure}}.$  Note that the $K$ side quantizers and the central quantizer are arranged periodically along the line $\mathbb{R}$. We only have to label the central points in a reference quantizer cell, which we will discuss in next subsection.


The Voronoi region $V(\lambda)$ of a point $\lambda$ in a quantizer $\mathcal{A}$ is defined to be
\begin{equation}
V(\lambda)=\{x|(x-\lambda)^2\leq (x-\tilde{\lambda})^2, \forall \tilde{\lambda}\in \mathcal{A}\}\mbox{,}
\label{equ:voronoi_region}
\end{equation} where the ties are broken in a systematic manner.
To describe the relation between the reference and central quantizers, we define a discrete Voronoi region associated with each $\lambda_r\in \mathcal{A}_r$~as
\begin{equation}
V_r(\lambda_r)=\{\lambda_c\in \mathcal{A}_c|(\lambda_c-\lambda_r)^2 < (\lambda_c-\lambda_r')^2, \forall \lambda_r'\in \mathcal{A}_r\}\mbox{.}
\label{equ:discrete_voronoi_region}
\end{equation}
The definitions of $\mathcal{A}_c$ and $\mathcal{A}_r$ guarantee that
the central points do not lie on the boundaries of the reference
cells. From (\ref{equ:central_quantizer}), the cardinality of
$V_r(\lambda_r)$ is $|V_r(\lambda_r)|=M$ for any
$\lambda_r\in\mathcal{A}_r$.

By studying (\ref{equ:side_quantizer}), the points of the side
quantizers are periodically and evenly distributed over the line
$\mathbb{R}$. Now we formulate this periodicity w.r.t. the reference
quantizer points. We assign a \textit{quantizer}-tuple
$\left(\mathcal{A}_0,\mathcal{A}_1,\ldots,\mathcal{A}_{K-1}\right)$ to
the element $0\in \mathcal{A}_r$ to describe the geometrical
relationship between the $K$ side quantizer and $\lambda_r=0$. Considering
the periodicity of side quantizers, the quantizer-tuple for a reference
quantizer point $\lambda_r=\zeta\cdot z$  is
\begin{equation}
\gamma(\zeta\cdot z)=\left(\mathcal{A}_{\{\textrm{mod}(z,K)\}},\ldots,\mathcal{A}_{\{\textrm{mod}(z+K-1,K)\}}\right)
\mbox{.}\label{equ:gamma_function}
\end{equation}
The $\gamma(\cdot)$ function in (\ref{equ:gamma_function}) exhibits
periodicity along with reference points $\zeta\cdot z$ with period
$K$. This geometrical property facilitates the design of optimal index
assignment and further the generation of the balanced
descriptions.

We now proceed with a definition of optimality of an index assignment. We first define a cost
\begin{equation}B(\lambda_c,\kappa)=\sum_{l\in
    \mathcal{L}^{(K,\kappa)}}(\lambda_c-\frac{1}{\kappa}\sum_{j=0}^{\kappa}\lambda_{l_j})^2\label{equ:labeling_evaluation}\end{equation}
to evaluate a labeling function for a particular central point $\lambda_c$ for the case of $\kappa$ descriptions received.
 Note that the defined cost is a geometrical measurement,
and is unrelated to channel conditions. The following theorem
decomposes the cost (\ref{equ:labeling_evaluation})
(see \cite{huang06Multiple}, \cite{xiaoqiang08MDC_lattice} for
details). The result applies to any dimensionality as long as the
$l_2$ norm is taken to measure the error.
\begin{theorem} \cite{huang06Multiple}
Suppose $\lambda_c \in \mathbb{R}^L$ is associated with a $K$-tuple $(\lambda_0,\lambda_1,\ldots,\lambda_{K-1})$, where $\lambda_i\in \mathbb{R}^{L}$, $i=0,\ldots,K-1$. Then for any $1\leq \kappa \leq K$:
\begin{eqnarray}&&\sum_{l\in \mathcal{L}^{(K,\kappa)}}\parallel \lambda_c-\frac{1}{\kappa}\sum_{j=1}^{\kappa}\lambda_{l_j}\parallel^2={K \choose \kappa}\Big[ \parallel \lambda_c-\bar{\lambda}\parallel^2\nonumber\\
&& +\frac{K-\kappa}{K\kappa(K-1)}\sum_{i=0}^{K-1}\parallel \lambda_i-\bar{\lambda} \parallel^2\Big]
\mbox{,}\label{equ:sideDistortion_de}
\end{eqnarray}
where $\parallel\cdot \parallel^{2}$ denotes the $l_2$ norm and $\bar{\lambda}$ denotes the centroid of the $K$-tuple, i.e. $\bar{\lambda}=\frac{1}{K}\sum_{j=0}^{K-1}\lambda_{j}$.
\label{theorem:sideDistortion_de}
\end{theorem}

The two terms on the right side of (\ref{equ:sideDistortion_de}) can
be interpreted geometrically. The first term measures the
squared distance (SD) between $\lambda_c$ and the centroid of a
$K$-tuple. The second term computes the sum of squared distances (SSD)
between the components and the centroid of the $K$-tuple, which
captures the geometrical structure of the $K$-tuple itself.

The optimal index assignment for the case of $\kappa$ ($\kappa<K$)
descriptions received is defined as
\begin{equation}
\alpha_{\mathrm{opt}}^{(K,\kappa)}=\min_{\alpha}\frac{1}{|\mathcal{A}_c|}\sum_{\lambda_c\in \mathcal{A}_c}B(\lambda_c,\kappa)\mbox{,}\label{equ:alpha_optimal}
\end{equation}
where $|\mathcal{A}_c|$ represents the cardinality of the central quantizer.
We note that $B(\lambda_c,\kappa)$ is a function of $\kappa$. By applying
(\ref{equ:sideDistortion_de}), (\ref{equ:alpha_optimal}) can be
rewritten as
\begin{align}
\alpha_{\mathrm{opt}}^{(K,\kappa)}& =& \min_{\alpha}\frac{1}{|\mathcal{A}_c|}\Big[\sum_{\lambda_c \in\mathcal{A}_c}(\lambda_c-\bar{\lambda}(\lambda_c))^2\nonumber\\
&&{ }+\frac{K-\kappa}{K\kappa(K-1)}\sum_{\lambda_c\in\mathcal{A}_c}\sum_{i=0}^{K-1}(a_i(\lambda_c)-\bar{\lambda}(\lambda_c))^2\Big],
\label{equ:alpha_optimal_2}
\end{align}
where
$\bar{\lambda}(\lambda_c)=\frac{1}{K}\sum_{i=0}^{K-1}\alpha_i(\lambda_c)$. The
right-hand side of equation
(\ref{equ:alpha_optimal_2}) consists of two parts, the SD
and SSD costs. We propose an index assignment that is optimal
with regard to any $\kappa<K$ in
(\ref{equ:alpha_optimal_2}). We show that the index assignment
minimizes both the SD and SSD costs simultaneously.

\subsection{Construction of Optimal Labeling Function}
Similarly to the two-channel case \cite{Vaishampayan93Multiple}, the
index assignment for the general $K$-channel case can be posed as a
geometrical minimization problem. We show that once the central points
in $V_r(0)$ are labeled, the rest can be obtained easily by exploring
the regularity of the central and side quantizers. The index
assignment is then shown to be optimal
w.r.t. (\ref{equ:alpha_optimal_2}).

From (\ref{equ:reference_side_relation}), it is known that the
centroid of any $K$-tuple used for index assignment is a reference
point. Thus, all the $K$-tuples can be grouped with respect to their
centroids or, equivalently, their reference points. Each reference point is
associated with many $K$-tuples. We evaluate those $K$-tuples with a
common centroid by measuring their SSD costs. We start with a
$K$-tuple with centroid $\lambda_r=0$. From (\ref{equ:side_quantizer}), any point of a side quantizer $\mathcal{A}_i$ is determined by its coordinate $x_i$. Thus, the SSD cost takes the form
\begin{equation}
\sum_{i=0}^{K-1}\lambda_i^2=\sum_{i=0}^{K-1}(K\zeta \cdot x_i+(2i-K+1)\frac{\zeta}{2})^2
\label{equ:IGC_cost}
\end{equation}
\begin{equation}
\textrm{subject to } \frac{1}{K}\sum_{i=0}^{K-1}\left[K\zeta \cdot x_i+(2i-K+1)\frac{\zeta}{2}\right]=0\mbox{.}\label{equ:centroid_constraint}
\end{equation}
The constraint is imposed to reflect the centroid condition.
Denote $X=[\begin{array}{cccc}x_0 & x_1 & \ldots & x_{K-1}\end{array}]^T$, representing a coordinate vector of a $K$-tuple. Eq. (\ref{equ:IGC_cost})--(\ref{equ:centroid_constraint}) can be further simplified as
\begin{equation}
J(X)=K^2\zeta^2 \parallel X+s\parallel^2\label{equ:J}
\end{equation}
\begin{equation}
\textrm{subject to } \sum_{i=0}^{K-1}x_i=0\mbox{,}
\label{equ:centroid_constraint2}\end{equation}
where
\begin{equation}s=\frac{1}{2K}[\begin{array}{cccc}-(K-1) & -K+3 & \ldots & K-1\end{array}]^T\mbox{.}\label{equ:translation_vector}\end{equation}
Take $X$ as a point in $K$-dimensional space. The constraint
(\ref{equ:centroid_constraint2}) forces the points to be in a
hyper-plane of dimensionality $K-1$. As each component
of $X$ only takes integer values, (\ref{equ:centroid_constraint2})
defines
an $A_{K-1}$ integral (the inner product of any two lattice vectors is an integer) lattice \cite{Conway98Book}. Thus, the
cost $J(X)$ in (\ref{equ:J}) can be interpreted as
measuring the squared distance of a point of a translated $A_{K-1}$
lattice from the origin regardless of the multiplying factor $K^2{\zeta}^2$. The translated lattice takes the form
\begin{equation}A_{K-1}+s\mbox{,}\end{equation}
where the vector $s$ is the translation. We denote the translated
lattice as $\check{A}_{K-1}$. Thus, one can associate all
$K$-tuple candidates that have common centroid $\lambda_r=0$ with
the points of $\check{A}_{K-1}$. The translated lattice
$\check{A}_{K-1}$ reveals the geometrical relation between the side
quantizers and $0\in A_r$. Note that each component $x_i$ of $X$ is the
coordinate of the side quantizer $\mathcal{A}_i$. This connects
$\lambda_r=0$ with the quantizer-tuple
$(\mathcal{A}_0,\mathcal{A}_1,\ldots,\mathcal{A}_{K-1})$ through $X$,
as is specified by the $\gamma(\cdot)$ function.

Next, let us consider the SSD cost of a $K$-tuple with a reference point $\zeta\cdot z$ as
its centroid, and further the geometrical
relation between $\zeta\cdot z$ and the side quantizers. Similarly to (\ref{equ:IGC_cost})-(\ref{equ:centroid_constraint}), the SSD cost can be expressed as
\begin{equation}
\sum_{k=0}^{K-1}(\lambda_i-\zeta\cdot z)^2=\sum_{i=0}^{K-1}(K\zeta \cdot x_i+(2i-K+1)\frac{\zeta}{2}-\zeta\cdot z)^2
\label{equ:IGC_cost_other}
\end{equation}
\begin{equation}
\textrm{subject to } \frac{1}{K}\sum_{i=0}^{K-1}\left[K\zeta \cdot
  x_i+(2i-K+1)\frac{\zeta}{2}\right]=\zeta\cdot
z\mbox{.}\label{equ:centroid_constraint_other}
\end{equation}
The above two expressions can also be simplified and characterized by the translated lattice $\check{A}_{K-1}$, which is described in the following Proposition.
\begin{proposition}
The SSD cost defined in
(\ref{equ:IGC_cost_other})-(\ref{equ:centroid_constraint_other}) has a
simplified form \begin{equation}J(\check{X}) \quad \textrm{subject to
  }
  \sum_{i=0}^{K-1}\check{x}_i=0\mbox{},\label{equ:J_other}\end{equation}
where
\begin{equation}
\check{X}=[\check{x}_i]_{i=0}^{K-1}=\left[\begin{array}{c}x_{\textrm{mod}(z,K)}-\lfloor \frac{z}{K}\rfloor\\
                                x_{\textrm{mod}(z+1,K)}-\lfloor \frac{z+1}{K}\rfloor\\
                                \vdots \\
                                x_{\textrm{mod}(z+K-1,K)-}\lfloor \frac{z+K-1}{K}\rfloor\\
                  \end{array}\right]\mbox{.}\label{equ:J_other_X}\end{equation}

\label{lemma:IGC_cost_form_other}
The operation $\lfloor x\rfloor$ takes the largest integer not exceeding $x$.
\end{proposition}

See Appendix \ref{proof:IGC_cost_form_other} for the proof.

Note that the $i$th component of $\check{X}$ in (\ref{equ:J_other_X})
is the coordinate of the side quantizer
$\mathcal{A}_{\textrm{mod}(z+i-1,K)}$. This again relates the
reference point $\zeta\cdot z$ with a quantizer-tuple through
$\check{X}$, confirming
(\ref{equ:gamma_function}). Eq. (\ref{equ:J_other}) shows that all
reference points are geometrically equivalent in terms of SSD cost as
$J(\check{X})$ always has the same structure. In other words, for every
$K$-tuple with centroid $\lambda_r=0$, there exists a $K$-tuple with a
centroid $\lambda_r=\zeta\cdot z$ that gives the same SSD cost.

The search procedure for good index assignment can be
performed in two steps. First, the central points within the
fundamental reference cell, $\{\lambda_c\in V_r(0)\}$, are considered
and labeled. The $K$-tuples exploited are enforced to have a common centroid $\lambda_r=0$. From (\ref{equ:J_other_X}), the labeling function
for any central point $\lambda_c+m\zeta \in \mathcal{A}_c$, $m\in
\mathbb{Z}$, $\lambda_c\in V_r(0)$, can then be obtained as
\begin{equation} \alpha_{\{\textrm{mod}(m+i,K)\}}(\lambda_c+m\zeta)=\alpha_i(\lambda_c)+\lfloor \frac{m+i}{K}\rfloor,\label{equ:general_labeling}\end{equation}
where $i=0,\ldots,K-1$. The operation defined in (\ref{equ:general_labeling}) guarantees that the $K$-tuples used to label the central points within $V_r(\lambda_r=\zeta\cdot z)$ have a centroid $\zeta\cdot z$. This ensures that no $K$-tuples are reused by performing (\ref{equ:general_labeling}). It is seen that the extension of the index
assignment from $V_r(0)$ to $V_r(\lambda_r)$ generally involves both
translation and permutation (changes of subscripts of a labeling function $\alpha_i$). The permutation
operation exhibits periodicity with period $K\zeta$. Thus, the
extension of the index assignment from
$\bigcup_{\lambda_r=0}^{(K-1)\zeta}V_r(\lambda_r)$ to
$\bigcup_{\lambda_r=0}^{(K-1)\zeta}V_r(\lambda_r+mK\zeta)$, $m\in
\mathbb{Z}$, only involves translation.

As described before, in order to label the $M$ central points in $V_r(0)$, we first constrain
$K$-tuple candidates to have centroid $\lambda_r=0$. This ensures that every central point contributes the corresponding minimum SD cost to (\ref{equ:alpha_optimal_2}). The selection of
good $K$-tuples for labeling can then be done by choosing points of
$\check{A}_{K-1}$. Specifically, we order the points of $\check{A}_{K-1}$ according to
their distances from origin. Denote the coordinate vector of the
$i$th point as $X_{i-1}$. The first $M$ coordinate vectors
$\{X_i,i=0,\ldots,M-1\}$ are then selected, which give the lowest $M$ SSD costs under the centroid constraint. We then define a bijective mapping \begin{equation}\beta:V_r(0)\rightarrow \{X_i,i=0,\ldots,M-1\}\label{equ:beta_function}\end{equation}
to relate the $K$-tuples to the
central points. Upon selecting a $\beta$ function, an index assignment is fully determined.
The cost in (\ref{equ:alpha_optimal_2}) incurred by the index assignment is given as
\begin{equation}
\frac{1}{M}\left[\sum_{\lambda_c\in V_r(0)}\lambda_c^2+\frac{K-\kappa}{K\kappa(K-1)}\sum_{i=0}^{M-1}J(X_i)\right]\mbox{.}
\label{equ:cost_opt_idx_assign}
\end{equation}
We now investigate the optimality of the index assignment.

\begin{theorem}
For the quantization structure defined by (\ref{equ:side_quantizer}) and (\ref{equ:central_quantizer}) and the averaging decoding strategy specified by (\ref{equ:X_reconstruction}), the index assignment specified by a $\beta$ function in (\ref{equ:beta_function}) and (\ref{equ:general_labeling}) is optimal w.r.t. (\ref{equ:alpha_optimal_2}) for any $\kappa<K$.
\label{theorem:idx_assign_opt}
\end{theorem}
\begin{IEEEproof}
The proof is trivial. First, the proposed index assignment guarantees
that the SD part of the costs in (\ref{equ:alpha_optimal_2}) is
minimized. Second, (\ref{equ:general_labeling}) and
(\ref{equ:beta_function}) imply that all the exploited $K$-tuples have
SSD costs not exceeding $J(X_{M-1})$. If unused $K$-tuples are taken
to replace exploited ones, the part of SSD costs in
(\ref{equ:alpha_optimal_2}) would obviously increase. On the other
hand, the part of SD costs would either increase or remain the same by switching the exploited
$K$-tuples of different central points. This shows that the proposed
index assignment is optimal for any $\kappa<K$.
\end{IEEEproof}

Note that a different $\beta$ function in (\ref{equ:beta_function})
does not affect (\ref{equ:cost_opt_idx_assign}). This implies that
there exist more than one optimal index assignment when $M>1$. The simplicity of
the proposed labeling function is due to the proper arrangement of the
central and side quantizers.

\subsection{Properties of the Index Assignment}
In this subsection, we first study the distribution of the points of
$\check{A}_{K-1}$. Specifically, its theta series \cite{Conway98Book}
is investigated for distortion evaluation. The issue of
generating balanced descriptions is then discussed. Finally, we consider the possibility of smooth adjustment of the redundancy among the descriptions.

To analyze the performance of the proposed index assignment, an
essential step is to study the properties of SSD costs. This
motivates us to look into the theta series of $\check{A}_{K-1}$, which
is defined as \cite{Conway98Book}
\begin{equation}\Theta_{\check{A}_{K-1}}(z)=\sum_{y\in \check{A}_{K-1}}q^{\parallel y \parallel^2}\mbox{,}\end{equation}
where $q=e^{\pi i z}$. It is seen that the theta series captures the
SSD costs of all the $K$-tuples that have a common centroid, rendering an infinite series. The theta series describes the distribution of the number of the translated lattice points with a common distance from origin. In principle, one can easily access the squared distances of the first $M$ points given the theta series, thus determining the part of SSD costs in (\ref{equ:cost_opt_idx_assign}). Further, if the theta series exhibits regularity (e.g. the number of points can be parameterized by their common distance from origin), the SSD costs of the first $M$ points can be computed easily. Let us denote by
$A^{\ast}_{K-1}$ the dual lattice (see Appendix
\ref{proof:theta_series_relation} or \cite{Conway98Book} for
definition) of $A_{K-1}$. The translation vector $s$ in
(\ref{equ:translation_vector}) is a deep hole \cite{Conway98Book} of
$A^{\ast}_{K-1}$. Informally speaking, deep holes of a lattice are the
points in the space that are farthest away from the lattice
points. The theta series of $\check{A}_{K-1}$ can be computed easily
with the aid of an $A^{\ast}_{K-1}$ lattice.
\begin{proposition}
Let $A^{\ast}_{K-1}(\textrm{hole})$ denote the translated lattice that is obtained by translating $A^{\ast}_{K-1}$ with one of its deep holes.
The theta series of $\check{A}_{K-1}$ is related to that of $A^{\ast}_{K-1}(\textrm{hole})$ through
\begin{equation}
\Theta_{\check{A}_{K-1}}(z)=\frac{1}{K}\Theta_{A^{\ast}_{K-1}(\textrm{hole})}(z)\mbox{.}
\label{equ:theta_relation}
\end{equation}
\label{proposition:theta_series_relation}
\end{proposition}
See the proof in Appendix \ref{proof:theta_series_relation}.

The theta series of lattices $A^{\ast}_{i}$, $i=1,2,3$, at their deep holes are well studied \cite{Conway98Book}, thus facilitating the analysis of $\check{A}_{i}$, $i=1,2,3$. For example, $\Theta_{\check{A}_{1}}$ takes the form
\begin{equation}\Theta_{\check{A}_{1}}=\sum_{m=0}^{\infty}q^{\frac{1}{2}(m+\frac{1}{2})^2}\mbox{.}\label{equ:Theta_K=1}\end{equation}
The expressions of the theta series
$\Theta_{A^{\ast}_{i}(\textrm{hole})}(z)$ for $i\geq 4$ remain to be
discovered.
Note that the theta series reveals the information about the number of
points with a particular Euclidean distance from the origin. One
application of the derived theta series is to verify the exploited
K-tuples in labeling the central points. The exploitation of the theta
series in computing the distortions $D^{(K,\kappa)}$, $i=1,\ldots,K$,
will be discussed later on.

Next we examine if the index assignment can produce balanced
descriptions. We study the cost incurred by a particular set of
received descriptions (corresponding to a $l\in
\mathcal{L}^{(K,\kappa)}$). Due to the periodicity of the index
assignment in (\ref{equ:general_labeling}), we measure the cost
\begin{equation}
\sum_{\lambda_c\in \bigcup_{\lambda_r=0}^{(K-1)\zeta}V_r(\lambda_r)}(\lambda_c-\frac{1}{\kappa}\sum_{j=1}^{\kappa}\lambda_{l_j})^2
\end{equation}
for a particular $l\in \mathcal{L}^{(K,\kappa)}$, $\kappa<K$.
For the two-channel case, the two components $\alpha_0$ and $\alpha_1$
for $\bigcup_{\lambda_r=0}^{\zeta}V_r(\lambda_r)$ are related~by
\begin{equation}
\left\{
\begin{array}{l}
\alpha_1(\lambda_c+\zeta)=\alpha_0(\lambda_c)+\zeta \\
\alpha_0(\lambda_c+\zeta)=\alpha_1(\lambda_c)+\zeta
\end{array}
\right.\mbox{,}\quad \lambda_c\in V_r(0)\mbox{.}
\nonumber
\end{equation}
This implies that the roles of the two side quantizers in index
assignment are changed symmetrically between $V_r(0)$ and
$V_r(\zeta)$. Thus, the two costs corresponding to the two elements in
$\mathcal{L}^{(2,1)}$ are the same, ensuring the generation of
balanced descriptions. Similar behavior is observed for the
three-channel case, which suggests that the index assignment for $K=3$
also
generates balanced descriptions. When $K>3$, other techniques are
required to produce balanced descriptions, e.g. time-sharing coding.

The parameter $M$ takes integer values. This implies that the adjustment
of the redundant information among descriptions cannot be performed
smoothly. To overcome this issue,
one can add an extra central cell with width smaller than $\zeta/M$
within each reference quantization cell \cite{Tian04MDC_scalar}.
The cell width can vary from 0 to $\zeta/M$ depending on the
redundancy needed. We will not analyze this method here.

\subsection{Index Assignment Matrix for the Two-description Case}
For the two-description case, the proposed index assignment can be
visualized by transforming it to an IA matrix. We provide a method to
parameterize the IA matrix. We show that the obtained IA matrix
is also optimal under a different
criterion~\cite{Berger-wolf99IndexAssignment}.

To construct an IA matrix from the index assignment, we index the
central quantizer points using its coordinate $y$. The coordinates of
the central points within $V_r(0)$ are
\begin{equation}
\{y=-\lfloor\frac{M}{2}\rfloor,-\lfloor\frac{M}{2}\rfloor+1,\ldots,-\lfloor\frac{M}{2}\rfloor+M-1\}\mbox{.}
\label{equ:z_fundamental_region}
\end{equation}
Similarly, we index the points of the two side quantizers
($\mathcal{A}_i$, $i=0,1$) by their coordinates $x_0$ and $x_1$. The
labeling function can then be simplified as $\alpha(y)=(x_0,x_1)$. The
inverse mapping is denoted as $\alpha^{-1}(x_0,x_1)=y$. A quantization
unit of side quantizer 0 is defined as
\begin{equation}
C_0(x_0)=\bigcup_{\alpha_0(y)=x_0}y\mbox{.}\nonumber
\end{equation}
Thus, a quantization unit $C_0(x_0)$ of side quantizer 0 is the
union of central quantizer points that map to $x_0$. A quantization
unit for side quantizer 1 is defined similarly.

\begin{figure*}[htb]
\centering
\begin{footnotesize}
  \psfrag{a}[c][c]{$\ldots$}
  \psfrag{b}[c][c]{$x_0$}
  \psfrag{c}[c][c]{$x_1$}

  \psfrag{e}[c][c]{(a)}
  \psfrag{f}[c][c]{(b)}
  \psfrag{d}[c][c]{(c)}

  \psfrag{m}[c][c]{(0)}
  \psfrag{g}[c][c]{(4)}
  \psfrag{h}[c][c]{(8)}
  \psfrag{i}[c][c]{(12)}

  \psfrag{l}[c][c]{(10)}
  \psfrag{k}[c][c]{(6)}
  \psfrag{j}[c][c]{(2)}

  \includegraphics[width=130mm]{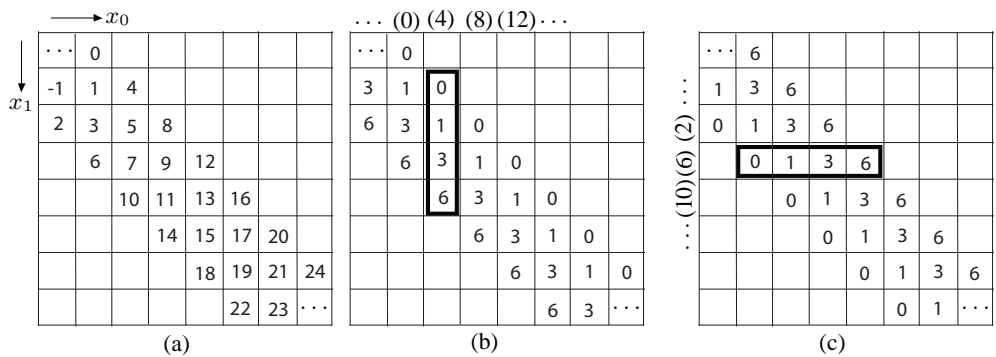}
\end{footnotesize}
\caption{(a) The IA matrix for $M=2$, where the bandwidth is 4. The elements in the matrix are the central cell coordinates. (b) The extracted patterns w.r.t. Side quantizer 0. (c) The extracted patterns w.r.t. side quantizer 1. The number in $(\cdot)$ describes the translation between a column(row) pattern and its original quantization unit.}
\label{fig:IA_matrix_N4}
\end{figure*}

To determine an IA matrix, we use the following $\beta$ function:
\begin{equation}
\beta(y=-\lfloor\frac{M}{2}\rfloor+i)=X_i,\quad i=0,\cdots,M-1.
\end{equation}
 In this situation, the redundancy index $N=2M$ becomes the bandwidth
 of the IA matrix. An
example for $M=2$ is given in Fig. \ref{fig:IA_matrix_N4} (a). It is seen from
the figure that any quantization unit from both the quantizers is in
fact a translated version of a fundamental pattern
$\{0,1,3,6\}$.  Fig. \ref{fig:IA_matrix_N4} (b) and (c)
display the translated quantization units of the two quantizers,
separately. Similar properties are observed for other $M$. As will
be shown below, the fundamental patterns for different $M$ can be
captured by an expression in terms of $M$:
\begin{eqnarray}
  P(M) = \{0\} \cup \{i(M-1)\}^{M-1}_{i=1} \cup
  \{(M-1)^{2}+M\}  \\ \nonumber \cup
  \{(M-1)^2+M+i(M+1)\}^{M-1}_{i=1}.\label{equ:P(M)}
\end{eqnarray}
where $M\geq 1$. An IA matrix can be built on $P(M)$.

We describe the scheme of how to extract the expression $P(M)$ from
the specified $\beta$ function. For the two-channel case, any 2-tuple
can also be related to a line segment connecting two side quantizer
points. The
length of a line segment is the associated SSD cost. This
implies that any line segment with length belonging to
\begin{equation}\{\zeta,3\zeta,\ldots,(2M-1)\zeta\}\label{equ:set_length}\end{equation}
 plays a role in the index
assignment. Without loss of generality, we consider the quantization
unit $C_0(0)$. There are $2M$ exploited line segments with one common endpoint $\lambda_0=-\zeta/2$ (equivalently $x_0=0$). In particular, any element in the set in (\ref{equ:set_length}) is associated with two exploited line segments, one having the left endpoint $-\zeta/2$ and the other having the right endpoint $-\zeta/2$. The corresponding 2-tuples are $\{(-\zeta/2,\zeta/2+2\zeta x_1)|x_1=-M,\ldots,M-1\}$, where $x_1$ is the coordinates of the points of side quantizer 1. By
computing the centroids of these 2-tuples, the coordinates of the
resulting reference points are related to $x_1$ by $z=x_1$. When $z$
increases from $-M$ to $-1$, the length of the associated line segment
decreases, implying
\begin{equation}\alpha^{-1}(0,i+1)-\alpha^{-1}(0,i)=M-1,\quad
  i=-M,\ldots,-2.\nonumber\end{equation}
The lengths of the line
segments when $z=-1$ and $z=0$ are equal to $\zeta$. This implies that
$\alpha^{-1}(0,0)-\alpha^{-1}(0,-1)=M$. Conversely, when $z$
increases from $0$ to $M-1$, the length of the associated line
segment increases, implying
\begin{equation}\alpha^{-1}(0,i+1)-\alpha^{-1}(0,i)=M+1,\quad
  i=0,\ldots,M-2.\nonumber\end{equation}
Combining the results from these three situations produces the pattern
$P(M)$ shown in (\ref{equ:P(M)}). Due to the symmetry of the side
quantization points, other quantization units $C_i(x_i)$, $i=0,1$,
give the same pattern.

We now present how to construct the IA matrix from $P(M)$. Without
loss of generality, assuming $C_0(0)$ is expressed by the pattern
$P(M)$ (usually, $C_0(0)$ is translated from $P(M)$). Then $C_0(x_0)$
is described by
\begin{equation}
P(M)+2Mx_0
\mbox{.}
\label{equ:IA_from_P(M)}
\end{equation}
The quantization units are arranged column-wise along the principle diagonal in a matrix, as shown in Fig. \ref{fig:IA_matrix_N4} (a). The constructed matrix implicitly determines the quantization units of side quantizer 1, which can be described as
$C_1(x_1):P(M)+2Mx_1+M$.

The IA matrix is systematically parameterized with the aid of
(\ref{equ:IA_from_P(M)}). This facilitates the generation of the IA
matrix. Due to the simplicity of $P(M)$, one can analytically investigate its asymptotic performance (e.g. \cite{Vaishampayan98Multiple}), or compute the side distortions and further derive optimal bandwidth in response to varying channel conditions.

An alternative principle of designing an IA matrix is to minimize the
so-called \emph{spread}, the difference between the minimum and the
maximum central indices within each quantization unit, subject to a
constant bandwidth \cite{Vaishampayan93Multiple}. A lower bound on the
spread is derived in \cite{Berger-wolf99IndexAssignment}, given as
$b(b-1)/2$ where $b$ is the bandwidth. It is immediate from
(\ref{equ:P(M)}) that each quantization unit achieves this bound. It
can also be shown that the IA matrix produced from any other $\beta$
function also achieves this bound. Thus, the derived IA matrix is also
optimal from a viewpoint of spread measurement.

\section{Evaluation of the Index Assignment}
In this section we evaluate the proposed index assignment. For brevity, we name the index assignment as \emph{$A_{K-1}$-based IA}. We mainly focus on the high-redundancy case, or equivalently, the index assignment with small $M$. This is because the high-redundancy case is more relevant in practice.

We consider encoding the Gaussian source $X\sim N(0,1)$, as an example. In the past, many theoretical results have been obtained for the Gaussian source. Thus, by choosing the Gaussian source, we are able to study the performance loss of the $A_{K-1}$-based IA.

One popular way to evaluate an index assignment is to analyze its performance under the high-rate assumption. By doing this, analytic expressions can often be obtained for approximating the corresponding side and central distortions. We point out that in most practical situations, both the irregularity of the index assignment and the high-rate approximation negatively affect the accuracy of the side distortion expressions (see \cite{vaishampayan01multiple,huang06Multiple,Klejsa08IndexAssignment,Ostergaard06MDC}). It may happen that different index assignments may have the same expressions for the side distortions, even though their real performance is different.

In the following, we show that the $A_{1}$-based IA (i.e., $K=2$) is regular. In other words, the accuracy of the side distortion expression is only affected by the high-rate approximation. This is due to the fact that the theta series of $A_{1}^{\ast}$ can be nicely parameterized (see (\ref{equ:Theta_K=1})). On the other hand, the theta series of $A_{K-1}^{\ast}$, $K>2$, take complicated forms. Thus, for the general $K$-channel case, we provide an expression to the side distortion by approximating the theta series.

Finally we will make an experimental comparison between the $A_{K-1}$-based IA and the index assignment of \cite{huang06Multiple}, which represents the state of the art in the literature. Our main focus is on the performance gain due to the side lattice translation.

\subsection{Performance Analysis for the Two-description Case}
We first consider the description rate for encoding the Gaussian source $X$. Assuming high-rate quantization, the per-channel rate $R$ can be shown to take the form (see \cite{vaishampayan01multiple}, \cite{Ostergaard06MDC})
\begin{equation}
R \approx \frac{1}{2}\log_2(2\pi e)-\log_2(2\zeta)\mbox{.}
\label{equ:side_rate}\end{equation}
It is seen that the rate $R$ is a function of the step size $\zeta$; if $\zeta$
is known the rate is independent of the index assignment. The minimum rate
required to transmit central indices is $R_c\approx
\frac{1}{2}\log_2(2\pi e)-\log_2(\frac{\zeta}{M})$. Thus, the rate overhead is
$2R-R_c\approx R_c-2\log_2(2M)$.

Next we study the central and side distortions, respectively. The central distortion $D_{(2,2)}$ is determined by the central quantizer, given as
\begin{equation}
D_{(2,2)} \approx \frac{\zeta^2}{12M^2}\mbox{.}
\label{equ:central_distortion}
\end{equation}
Note that the descriptions for the two-channel case are balanced, resulting in identical side distortions. The side distortion $D_{(2,1)}$ can be approximated as
\begin{eqnarray}
D_{(2,1)}\hspace{-2mm}&=&\hspace{-2mm}\sum_{\lambda_c \in \mathcal{A}_c}\int_{V(\lambda_c)}f_{X}(x)(x- \alpha_0(\lambda_c))^2dx \nonumber \\
\hspace{-2mm}&\approx &\hspace{-2mm} D_{(2,2)}+\frac{1}{2M}\sum_{\lambda_c\in V_r(0)}\big((\lambda_c-\alpha_0(\lambda_c))^2 \nonumber \\ &&+(\lambda_c-\alpha_1(\lambda_c))^2\big) \nonumber \\
\hspace{-2mm}&=&\hspace{-2mm}D_{(2,2)}+\frac{1}{M}\sum_{\lambda_c\in V_r(0)}\lambda_c^2+\frac{1}{2M}\sum_{i=0}^{1}\alpha_i(\lambda_c)^2
\label{equ:Ds} \mbox{.}
\end{eqnarray}
The part of SD costs $\frac{1}{M}\sum_{\lambda_c\in V_r(0)}\lambda_c^2$ has a simple expression
\begin{equation}
\frac{1}{M}\sum_{\lambda_c\in V_r(0)}\lambda_c^2=(1-\frac{1}{M^2})\frac{\zeta^2}{12}\mbox{.}\label{equ:side_distortion_term2}
\end{equation}
The part of SSD costs in (\ref{equ:Ds}) are related to $J(X)$~by
\begin{equation}
\sum_{\lambda_c\in V_r(0)}\sum_{i=0}^{1}\alpha_i(\lambda_c)^2=\sum_{i=0}^{M-1}J(X_i)\mbox{.}
\label{equ:side_distortion_term3}
\end{equation}
Combining (\ref{equ:J}), (\ref{equ:Theta_K=1}) and
(\ref{equ:side_distortion_term3}) yields
\begin{eqnarray}
\sum_{i=0}^{M-1}J(X_i) &=& 4\zeta^2\sum_{m=0}^{M-1}\frac{1}{2}(m+\frac{1}{2})^2 \nonumber \\
&=& \zeta^2(\frac{2M^3}{3}-\frac{M}{6}) \mbox{.}
\label{equ:side_distortion_term3_K=2}\end{eqnarray}
Finally, inserting (\ref{equ:central_distortion}), (\ref{equ:side_distortion_term2}) and (\ref{equ:side_distortion_term3_K=2}) into (\ref{equ:Ds}) produces
\begin{equation}D_{(2,1)}\approx\frac{M^2\zeta^2}{3}\mbox{.}\label{equ:Ds_final}\end{equation}

We now study the performance loss of the $A_1$-based IA. Computing the product of the side and central distortions from (\ref{equ:central_distortion}) and (\ref{equ:Ds_final}) gives $D^{(2,1)}D^{(2,2)}\approx\frac{\zeta^4}{3\cdot 12}$. This suggests that the product is unrelated to the parameter $M$ which serves as a trade-off factor between side and central distortions. As $\zeta$ is determined by the rate $R$, the product can be rewritten as a function of $R$, as given as
\begin{equation}
D_{(2,1)}D_{(2,2)}\approx\frac{1}{4}\frac{(2\pi e)^2}{144}2^{-4R}\mbox{.}\label{equ:distortion_product}
\end{equation}
The approximation (\ref{equ:distortion_product}) is valid for any ratio of central and side distortions as long as high-rate assumption holds.
The work in \cite{Vaishampayan98Multiple} derived an approximation to the Gaussian MD rate-distortion function, which is given by
\begin{equation}
\bar{D}_{(2,2)}\bar{D}_{(2,1)}\approx \frac{1}{4}2^{-2R}, \quad R\rightarrow \infty.\label{equ:RD_opt_two}
\end{equation}
It is seen that the gap between (\ref{equ:distortion_product}) and (\ref{equ:RD_opt_two}) is constant, which is characterized by $\frac{(2\pi e)^2}{144}$.

\subsection{Performance Analysis for the $K$-description Case}
Consider high-rate quantization for the $K$-description case. The rate $R$ remains the same as  (\ref{equ:side_rate}). Similarly, the central distortion $D_{(K,K)}$ takes the same form as $D_{(2,2)}$, i.e.,
\begin{equation}
D_{(K,K)}\approx\frac{\zeta^2}{12M^2}.\label{equ:central_dis_gen}
\end{equation}
Under the high-rate assumption, the side distortion $D_{(K,\kappa)}$ is approximated as
\begin{eqnarray}
D_{(K,\kappa)}&\approx &  D_{(K,K)}+\frac{1}{M}\sum_{\lambda_c\in V_r(0)}\lambda_c^2\nonumber \\
&&+\frac{1}{M}\frac{K-\kappa}{K\kappa(K-1)}\sum_{\lambda_c\in V_r(0)}\sum_{i=0}^{K-1}\alpha_i(\lambda_c)^2. \label{equ:sideDis_gen}
\end{eqnarray}
The last term in (\ref{equ:sideDis_gen}) is closely related to the theta series $\Theta_{\breve{A}_{K-1}}$. By using the same analysis as in \cite{Vaishampayan94MDC}, the side distortion can be approximated as
\begin{equation}
D^{(K,k)}\approx\frac{\zeta^2}{12}\Big[1+\frac{K(K-k)}{k}K^{\frac{1}{K-1}}\frac{G(S_{K-1})}{G(S_1)}M^{\frac{2}{K-1}}\Big]\mbox{,}
\label{equ:D^(K,k)_2}
\end{equation}
where $G(S_{i})$ is the normalized second moment of a sphere in the $i$ dimensional space. Note the derivation from (\ref{equ:sideDis_gen}) to (\ref{equ:D^(K,k)_2}) involves an approximation of the theta series, which is not necessary for the two-channel case.

Similarly to the two-channel case, some theoretical results for $K$-channel MD have been obtained in the past. When only the individual side distortions (corresponds to $\kappa=1$) and central distortion are of primary concern in an MD system, the symmetric Gaussian MD rate-distortion function has been derived in \cite{Wang07VectorMDC}. The work of \cite{Zhang2011} analyzed the rate-distortion function and derived a simple approximation. Specifically, the MD rate-distortion can be approximated as \cite{Zhang2011}
\begin{equation}
\bar{D}_{(K,K)}\left(\bar{D}_{(K,1)}\right)^{K-1}\approx (K-1)^{K-1}K^{-K}2^{-2KR}, \label{equ:RD_opt_gen}
\end{equation}
when $R\rightarrow \infty$. In the next subsection for experimental comparison, we will take the approximation (\ref{equ:RD_opt_gen}) as a reference, which we denote as \emph{R-D opt.}.

\subsection{Experimental Evaluation}
It is known that the construction of translated scalar lattices would bring the staggered gain to an MD system. Thus, in principle, the $A_{K-1}$-based IA would outperform the index assignment of \cite{huang06Multiple} in the same quantization space. In this subsection, we evaluate the performance gain of the $A_{K-1}$-based IA over that of \cite{huang06Multiple}.

\begin{figure*}[ht]
\centering
\subfigure[$K=2$]{
\psfrag{a}[c][c]{\footnotesize{$D_{(2,2)}$ (dB)}}
\psfrag{d}[c][c]{\footnotesize{$D_{(2,1)}$ (dB)}}
\psfrag{q}[l][l]{\footnotesize{IA of \cite{huang06Multiple}}}
\psfrag{w}[l][l]{\footnotesize{$A_1$-based IA}}
\psfrag{e}[l][l]{\footnotesize{HR-approx.}}
\psfrag{r}[l][l]{\footnotesize{R-D opt. }}
\includegraphics[scale=0.600]{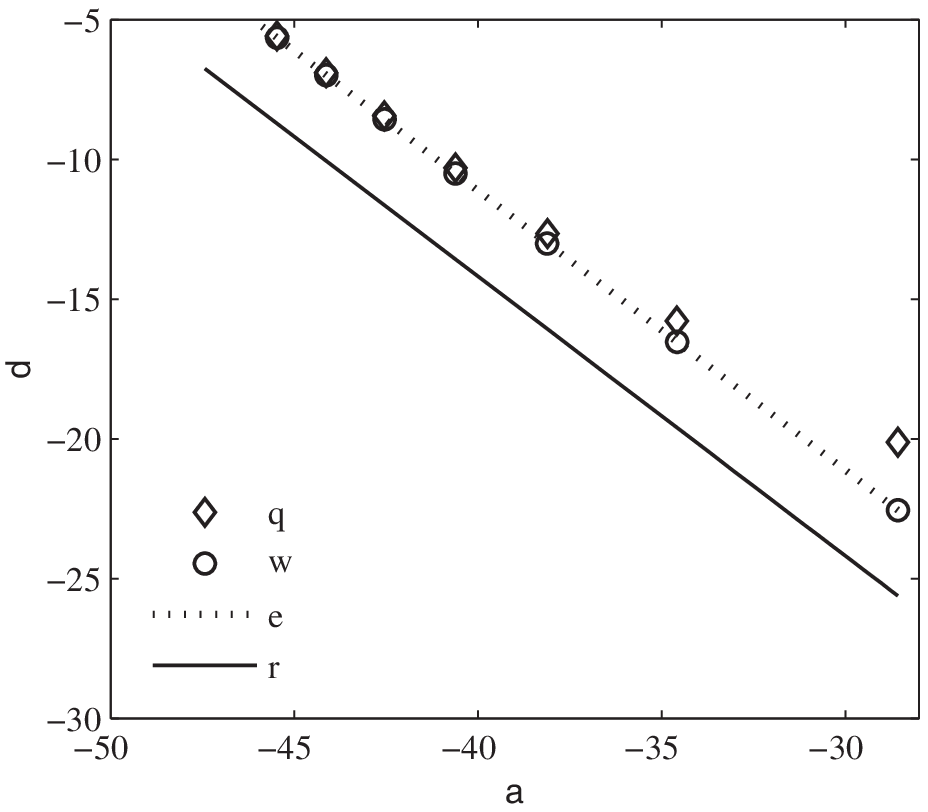}}
\subfigure[$K=3$]{
\psfrag{b}[c][c]{\footnotesize{$D_{(3,3)}$ (dB)}}
\psfrag{e}[c][c]{\footnotesize{$D_{(3,1)}$ (dB)}}
\psfrag{t}[l][l]{\footnotesize{IA of  \cite{huang06Multiple}}}
\psfrag{y}[l][l]{\footnotesize{$A_2$-based IA}}
\psfrag{u}[l][l]{\footnotesize{HR-approx.}}
\psfrag{i}[l][l]{\footnotesize{R-D opt.}}
\includegraphics[scale=0.600]{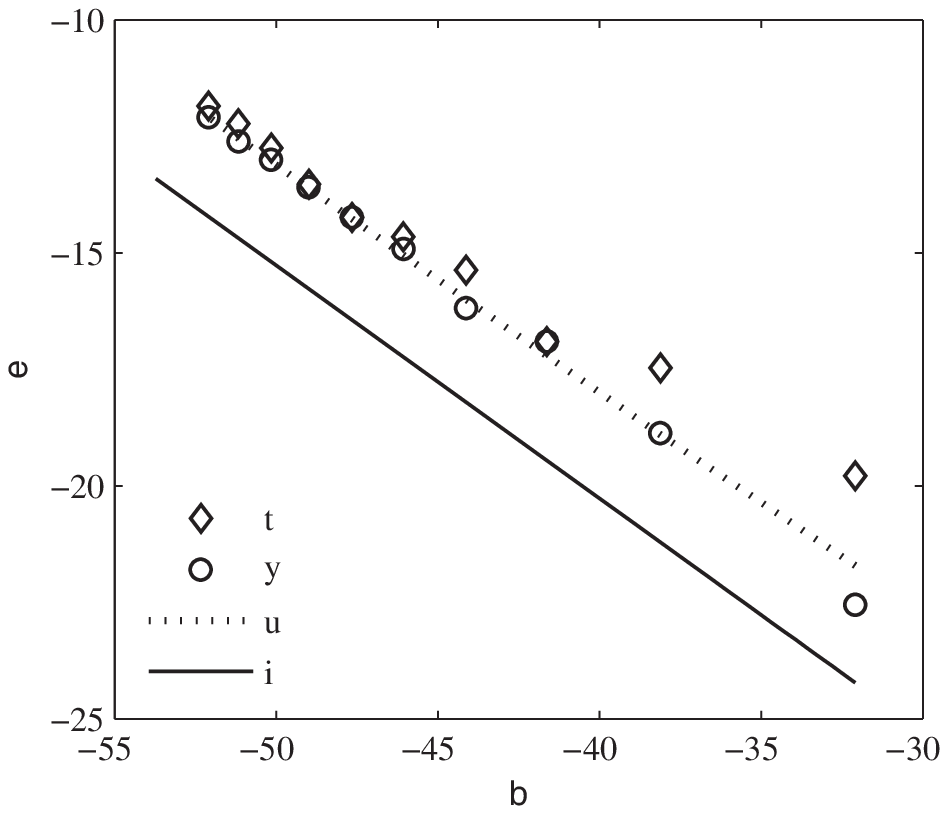}}
\subfigure[$K=4$]{
\psfrag{c}[c][c]{\footnotesize{$D_{(4,4)}$ (dB)}}
\psfrag{f}[c][c]{\footnotesize{$D_{(4,1)}$ (dB)}}
\psfrag{o}[l][l]{\footnotesize{IA of  \cite{huang06Multiple}}}
\psfrag{p}[l][l]{\footnotesize{$A_3$-based IA}}
\psfrag{s}[l][l]{\footnotesize{HR-approx.}}
\psfrag{x}[l][l]{\footnotesize{R-D opt.}}
\includegraphics[scale=0.600]{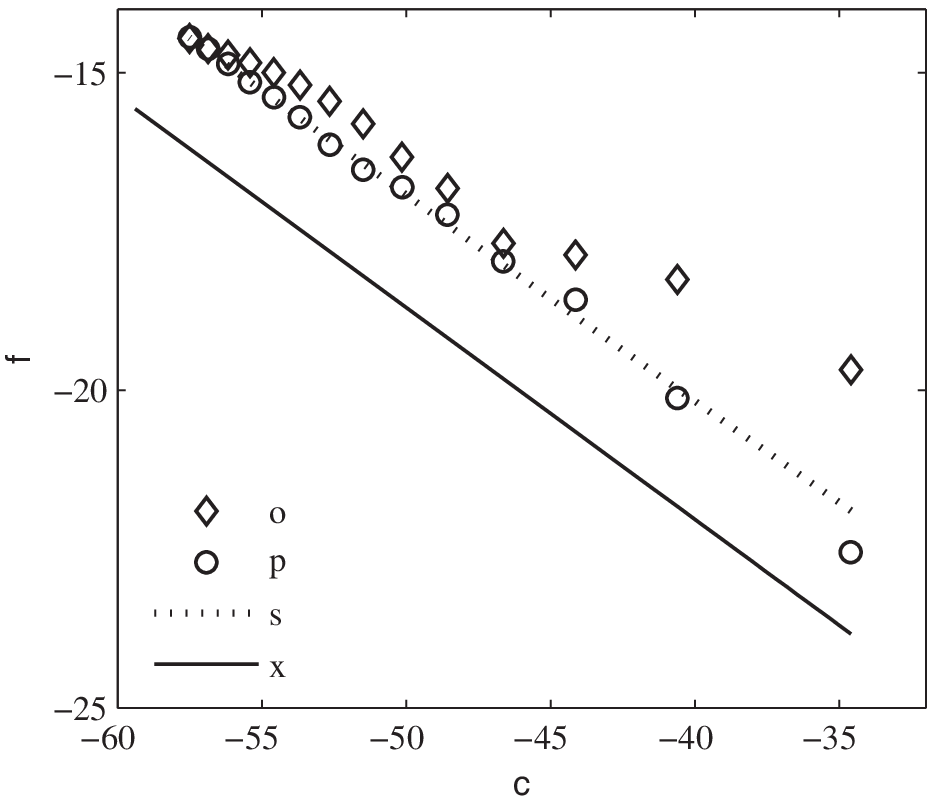}}
\caption{A trade-off between $D^{(K,K)}$ and $D^{(K,1)}$ for the index assignment based on the $A_{K-1}$ lattice and the index assignment of $\cite{huang06Multiple}$ for different values of $K$ and $R=4$ (bits). The parameter $M$ in the three experiments takes values of $M=\{1,2,\ldots,7\}$, $M=\{1,2,\ldots,10\}$ and $M=\{1,2,\ldots,14\}$, respectively.}
\label{fig:peform_compare}
\end{figure*}

In order to focus on the index assignments, we consider high-rate quantization. Since we are interested in high-redundancy case (small $M$), the high-rate assumption is reasonable when the per-description rate $R$ is large. When $K=2$, we tested the central distortion $D_{(2,2)}$ in (\ref{equ:central_distortion}) versus the side distortion $D_{(2,1)}$ in (\ref{equ:Ds_final}). For multi-channel case, we tested the central distortion $D_{(K,K)}$ in (\ref{equ:central_dis_gen}) versus the individual side distortion $D_{(K,1)}$ in (\ref{equ:sideDis_gen}) for $K=3,4$, respectively. We used the same operating rates for the index assignment of \cite{huang06Multiple}. The experimental results are presented in Fig.~\ref{fig:peform_compare}, one subplot for each number of descriptions. For comparison, we also plot the individual side distortion approximation (\ref{equ:Ds_final}) or (\ref{equ:D^(K,k)_2}) in respective subplots.

It is seen from the figure that when $M$ is small, the performance gain of the $A_{K-1}$-based IA over that of  \cite{huang06Multiple} due to lattice-translation is considerable. For the special case that $M=1$ in the three experiments, the gain is above $2$ dB. Further, the performance gain increases along with $K$.  This observation suggests that for general $K$-description quantizer, it is worth to use the structure of translated lattices.

Finally, we conclude from the figure that the side distortion approximation $D_{(K,1)}$ in (\ref{equ:D^(K,k)_2}) is accurate even for small $M$. The real performance of the $A_{K-1}$-based IA fluctuates closely around the derived approximation (\emph{HR-Approx.} in Fig.~\ref{fig:peform_compare}). This suggests that in practice, one can use the approximation (\ref{equ:D^(K,k)_2}) to configure the MD system (finding the optimal $M$) for a particular channel condition.

\section{Conclusion}

\par We conclude that the proposed index assignment provides a performance gain that is significant from the point of view of practical MDC applications that operate with a non-vanishing redundancy among the descriptions. The labeling function $\alpha$ based on the $A_{K-1}$ lattice exploits the staggered gain. It also leads to low operational complexity of scalar MDC as virtually the index assignment can be computed at hand. The use of the $A_{K-1}$ lattice facilitates analysis of the rate-distortion performance of a $K$-description scalar quantizer and analytic derivation of distortion for any description loss scenario. As a result, the obtained MDC scheme can be analytically optimized with respect to channel conditions enabling an instantaneous re-optimization of the scheme.
\par Our results demonstrate that for the considered optimality criterion, there exist many index assignments schemes that are equivalent in terms of their performance. In principle any scheme from the group of the optimal index assignment schemes may be selected. However, certain schemes may have properties that make them particularly attractive. In particular, the proposed index assignment for the two-description case generates a periodic pattern of side quantization cells. This may lead to a low operational complexity of the resulting two-description quantizer or facilitate further extensions such as dithering (e.g. \cite{Frank-Dayan02MDC}).

\appendices


\section{Proof of Lemma \ref{lemma:IGC_cost_form_other}}
\label{proof:IGC_cost_form_other}
\begin{IEEEproof}
To prove the proposition, we first prove that (\ref{equ:IGC_cost_other})-(\ref{equ:centroid_constraint_other}) take the form
\begin{equation}
\sum_{j=0}^{K-1}\left[K\zeta \cdot (x_{\textrm{mod}(j+z,K)}-\lfloor\frac{j+z}{K}\rfloor)+(2j-K+1)\frac{\zeta}{2}\right]^2
\label{equ:IGC_cost_other_2}\end{equation}
\begin{eqnarray}
\textrm{subject to }&& \frac{1}{K}\sum_{j=0}^{K-1}\left[K\zeta \cdot (x_{\textrm{mod}(j+z,K)}-\lfloor\frac{j+z}{K}\rfloor)\right.\nonumber\\
&&\left.+(2j-K+1)\frac{\zeta}{2}\right]=0\mbox{.}
\label{equ:centroid_constraint_other_2}
\end{eqnarray} We apply the induction argument to prove this.

It is obvious that
(\ref{equ:IGC_cost_other_2})-(\ref{equ:centroid_constraint_other_2})
hold for $z=0$. Next, we study the case that $z=1$. Eq. (\ref{equ:IGC_cost_other})
can thus be rewritten as
\begin{eqnarray}
\hspace*{-2mm}&&\hspace*{-2mm}\frac{1}{K}\sum_{i=0}^{K-1}\left[K\zeta \cdot x_i+(2i-1-K+1)\frac{\zeta}{2}-\zeta\right]^2\nonumber \\
\hspace*{-2mm}&=&\hspace*{-2mm}\frac{1}{K}\sum_{i=0}^{K-1}\left[K\zeta \cdot x_i+(2(i-1)-K+1)\frac{\zeta}{2}\right]^2\nonumber \\
\hspace*{-2mm}&=&\hspace*{-2mm}\frac{1}{K}\sum_{j=0}^{K-2}\left[K\zeta \cdot x_{j+1}+(2j-K+1)\frac{\zeta}{2}\right]^2\nonumber \\
\hspace*{-2mm}&&\hspace*{-2mm}+\frac{1}{K}\left[K\zeta(x_{\textrm{mod}(1+K-1,K)}-\lfloor\frac{1+K-1}{K}\rfloor)+(K-1)\frac{\zeta}{2}\right]^2 \nonumber\\
\hspace*{-2mm}&=&\hspace*{-2mm}\frac{1}{K}\sum_{j=0}^{K-1}\left[K\zeta \cdot (x_{\textrm{mod}(j+1,K)}-\lfloor\frac{j+1}{K}\rfloor)+(2j-K+1)\frac{\zeta}{2}\right]^2\nonumber
\mbox{,}\end{eqnarray}
which is consistent with (\ref{equ:IGC_cost_other_2}). Eq. (\ref{equ:centroid_constraint_other}) for $z=1$ is expressed as
\begin{eqnarray}
&& \frac{1}{K}\sum_{i=0}^{K-1}\left[K\zeta \cdot x_i+(2i-K+1)\frac{\zeta}{2}\right]=\zeta \nonumber\\
&& \frac{1}{K}\sum_{i=0}^{K-1}\left[K\zeta \cdot x_i+(2i-K+1)\frac{\zeta}{2}-\zeta\right]=0\mbox{.}\nonumber
\end{eqnarray}
Following the same derivation as for that of (\ref{equ:IGC_cost_other_2}),
the above equation can be rewritten in the form of
(\ref{equ:centroid_constraint_other_2}). Next we assume the two
equations hold for $z=k$. Similarly, we can derive the expressions for
$z=k+1$ based on those for $z=k$. As for $z\in \mathbb{Z}^{-}$, the
argument is the same. This implies that
(\ref{equ:IGC_cost_other_2})-(\ref{equ:centroid_constraint_other_2})
hold for any $z\in \mathbb{Z}$.

Note that (\ref{equ:IGC_cost_other_2})--(\ref{equ:centroid_constraint_other_2}) have the same structures as
(\ref{equ:IGC_cost})-(\ref{equ:centroid_constraint}). By proper
variable replacement, the equivalent form of
(\ref{equ:IGC_cost_other_2})
is $J_{\textrm{IGC}}(\check{X})$, where $\check{X}$ is as given by
(\ref{equ:J_other_X}). This completes the proof.
\end{IEEEproof}

\section{Proof of Proposition \ref{proposition:theta_series_relation}}
\label{proof:theta_series_relation}
\begin{IEEEproof}
The dual lattice $A^{\ast}_{K-1}$ is defined as \cite{Conway98Book}
\begin{equation}A^{\ast}_{K-1}=\bigcup_{i=0}^{K-1}([i]+A_{K-1})\mbox{,}\end{equation}
where
\begin{equation}[i]=\left(\frac{i}{K},\cdots,\frac{i}{K},\frac{i-K}{K},\cdots,\frac{i-K}{K}\right)^T,\end{equation}
with $K-i$ components equal to $i/K$ and $i$ components equal to
$(i-K)/K$, are called \textit{glue vectors}.

Let $e=(1,1,\cdots,1)^T$, which has $K$ components. We rewrite the translating vector $s\textrm{ as}$
\begin{equation}s=\frac{1-K}{2K}e+\frac{1}{K}(0,1,\cdots,K-1)^T\label{equ:translation_vector_2}\mbox{.}\end{equation}
Next we define a shifting operation on a vector as
\begin{equation}SH([a_0,\cdots,a_{K-2},a_{K-1}]^T)=[a_1,\cdots,a_{K-1},a_0]^T\mbox{.}\end{equation}
Let $s_{[i]}=SH^{i}(s)$ denote the vector that is obtained by performing $i$ times the shifting operations on $s$. With the aid of (\ref{equ:translation_vector_2}), we can easily conclude that
\begin{equation}s_{[i]}=s+[i]\mbox{.}\nonumber\end{equation} This
shows that the theta series of $\check{A}_{K-1}$ is identical to
those of the translated lattices $A_{K-1}+s+[i]$, $1\leq i\leq
K-1$. Thus, we have that
\begin{equation}\Theta_{\check{A}_{K-1}}(z)=\frac{1}{K}\Theta_{\{A^{\ast}_{K-1}+s\}}(z)\mbox{.}\end{equation}
As $s$ is a deep hole of $A^{\ast}_{K-1}$, this proves the proposition.
\end{IEEEproof}

\bibliographystyle{IEEEtran}

%

%
%
%




\end{document}